\documentclass[aps,twocolumn,prb,floatfix,citeautoscript,showpacs]{revtex4}
\usepackage[dvips]{graphicx}
\usepackage{bm}
\usepackage[normalem]{ulem}
\usepackage{amsmath,amssymb}

\newcommand{\ab}{{\alpha\beta}}
\newcommand{\dab}{{\delta_\ab}}

\newcommand{\Rij}{{\bm{R}_{ij}}}
\newcommand{\Rijn}{{\bm{R}_{ij}+\bm{n}}}
\newcommand{\Gunit}{{\frac{2\pi}{a_0}}}
\newcommand{\eInf}{{\epsilon_\infty}}
\newcommand{\La}{{L_\alpha}}

\begin{document}
\title{Fast Molecular-Dynamics Simulation for Ferroelectric Thin-Film Capacitors
Using a First-Principles Effective Hamiltonian}

\author{Takeshi Nishimatsu$^{1,2}$}
\author{Umesh V. Waghmare$^{3}$}
\author{Yoshiyuki Kawazoe$^{2}$}
\author{David Vanderbilt$^{1}$}

\affiliation{
$^{1}$Department of Physics and Astronomy, Rutgers University,
136 Frelinghuysen Road, Piscataway, NJ 08544-8019\\
$^{2}$Institute for Materials Research (IMR), Tohoku University, Sendai 980-8577, Japan\\
$^{3}$Theoretical Sciences Unit, Jawaharlal Nehru Centre for Advanced Scientific Research (JNCASR),
Jakkur, Bangalore, 560 064, India
}

\begin{abstract}
A newly developed fast molecular-dynamics method is applied to BaTiO$_3$
ferroelectric thin-film capacitors with short-circuited electrodes
or under applied voltage.
The molecular-dynamics simulations based on a first-principles effective
Hamiltonian clarify that dead layers (or passive layers) between ferroelectrics
and electrodes markedly affect the properties of capacitors,
and predict that the system is unable to hop between a
uniformly polarized ferroelectric structure and
a striped ferroelectric domain structure at low temperatures.
Simulations of hysteresis loops of thin-film capacitors are also performed,
and their dependence on film thickness, epitaxial constraints,
and electrodes are discussed.
\end{abstract}

\pacs{77.80.Dj, 77.80.Fm, 64.70.Nd}


\maketitle

\section{Introduction}
Ferroelectric thin films are beginning to see wide-ranging applications,
for example in multilayer capacitors,
nonvolatile FeRAMs~\cite{Scott:Ferroelectric:Memories:2000},
and nanoactuators.
There is strong pressure to reduce the sizes of such
thin-film structures.
In recent years,
the preparation of oxide thin films by low-temperature
non-equilibrium techniques such as
molecular beam epitaxy and pulsed-laser deposition
have attracted a great deal of attention, as they enable finely
controlled growth of epitaxial thin
films~\cite{Choi:B:L:S:S:U:R:C:P:G:C:S:E:Science:306:p1005-1009:2004}.

It is well known that the properties of ferroelectric capacitors are
highly influenced by the properties of the interface between
the ferroelectrics and the electrodes.  For example,
the fatigue of ferroelectric capacitors is associated with the appearance of
dead layers (or passive layers) near the electrodes,~\cite{Drougard:L:JAP:30:p1663-1668:1959,
Miller:N:S:R:D:JAP:68:p6463-6471:1990,
Lemanov:Yarmarkin:PhysSolidState:38:1996,
Jin:Z:JAP:92:p4594-4598:2002}
and imperfect electrodes cannot fully screen the polarization
of ferroelectrics~\cite{MEHTA:S:J:JAP:44:p3379-3385:1973,Dawber:C:L:S:JPhys-CondesMatter:15:pL393-L398:2003},
leading to a finite depolarization field in the ferroelectric film.
However, the nanosize effects and temperature dependences of
ferroelectric capacitor hysteresis, polarization switching, and dynamics
of domain wall motion remain poorly known.
Experimentally, {\em in situ} observations are difficult.
Theoretically, the long-range Coulomb interaction limits the size
and time of molecular-dynamics (MD) simulations, and
it has been unclear how to include surface effects and
depolarization fields caused by interface structures.

In 1994, King-Smith and Vanderbilt studied the total-energy surface
for zone-center distortions of perovskite-type ferroelectric oxides
$AB$O$_3$ ($A$ is a monovalent or divalent cation and
$B$ is a penta- or tetravalent metal)
at zero temperature
using first-principles calculations with ultrasoft-pseudopotentials and
a plane-wave basis set.~\cite{King-Smith:V:1994}
Starting from the full symmetric cubic perovskite structure,
they define the displacements $v_{\alpha}^{\tau}$
of atoms $\tau$ (=$A$, $B$, O$_{\rm I}$, O$_{\rm II}$, O$_{\rm III}$)
in the Cartesian directions $\alpha (=x,y,z)$
along the
$\Gamma_{15}$ soft-mode normalized
direction vectors $\bm{\xi}_\alpha$ as
\begin{equation}
  \label{eq:Eigenvector}
\bm{v}_\alpha=
\left(
  \begin{array}{c}
    v_\alpha^A         \\
    v_\alpha^B         \\
    v_\alpha^{\rm O_{\rm I}} \\
    v_\alpha^{\rm O_{\rm II}} \\
    v_\alpha^{\rm O_{\rm III}}
  \end{array}
\right)
= u_\alpha\bm{\xi}_\alpha
= u_\alpha
  \left(
    \begin{array}{c}
      \xi_\alpha^A         \\
      \xi_\alpha^B         \\
      \xi_\alpha^{\rm O_{\rm I}} \\
      \xi_\alpha^{\rm O_{\rm II}} \\
      \xi_\alpha^{\rm O_{\rm III}}
    \end{array}
  \right)\ ,
\end{equation}
with the scalar soft-mode amplitude $u_\alpha$.
Under the condition that the strain components
$\eta_i$~($i=1,\ \dots,\ 6$; Voigt notation; $\eta_1=e_{11}$, $\eta_4=e_{23}$)
minimize the total energy for each $\bm{u}=(u_x, u_y, u_z)$,
they expressed the total energy as
\begin{equation}
  \label{eq:King-Smith-and-Vanderbilt}
  E^{\rm tot}
  = E^{0}
  + \kappa u^2
  + \alpha ' u^4
  + \gamma ' (u_x^2 u_y^2 +
              u_y^2 u_z^2 +
              u_z^2 u_x^2)\ ,
\end{equation}
where $u^2 = u_x^2 + u_y^2 + u_z^2$,
$E^{0}$ is the total energy for the cubic structure,
$\kappa$ is half the eigenvalue of the soft mode,
and $\alpha '$ and $\gamma '$ are the constants determined from
coupling constants between
atomic displacements and strains.
Their expression properly describes the coupling of
polar atomic-displacement and strain degrees of freedom.

In 1994-1997, Zhong, Vanderbilt,
and Rabe~\cite{Zhong:V:R:1994,Zhong:V:R:PRB:v52:p6301:1995}
and Waghmare and Rabe~\cite{Waghmare:R:1997PRB}
expanded Eq.~(\ref{eq:King-Smith-and-Vanderbilt})
from a mean-field framework to a local-mode framework,
replacing ${\bm u}$ by $\{\bm{u}\}$,
where the braces $\{\}$ denote a {\em set} of $\bm{u}$ in a
simulation supercell, as
\begin{multline}
  \label{eq:MC:Hamiltonian}
  E^{\rm tot} =
    V^{\rm self}(\{\bm{u}\}) + V^{\rm dpl}(\{\bm{u}\})+V^{\rm short}(\{\bm{u}\})\\
  + V^{\rm elas}(\eta_1,\cdots\!,\eta_6)   +   V^{\rm int}(\{\bm{u}\}, \eta_1,\cdots\!,\eta_6)~.
\end{multline}
Here
$V^{\rm self}$, $V^{\rm dpl}$, $V^{\rm short}$, $V^{\rm elas}$, and $V^{\rm int}$ are
a local-mode self-energy,
a long-range dipole-dipole interaction,
a short-range interaction between soft-modes,
an elastic energy,
and an interaction between the local modes and local strain, respectively.
They employed Eq.~(\ref{eq:MC:Hamiltonian})
as an effective Hamiltonian for $\{\bm{u}\}$ in
the supercell,
performed Monte-Carlo simulations,
and
demonstrated the ability to describe the phase transitions of bulk ferroelectrics.
The coarse-graining that reduces the
15-dimensional atomic displacement vector $v_{\alpha}^{\tau}$
to a 3-dimensional local soft-mode amplitude vector $u_\alpha$
in each unit cell was shown to be a good approximation.
However, the computation of $V^{\rm dpl}$ was still time-consuming,
owing to the long-range Coulomb interaction, thus
limiting system size and simulation time that could be handled in
practical simulations.

In 2003, Waghmare, Cockayne, and Burton introduced a technique
to decrease the computational time
for $V^{\rm dpl}$ (or forces exerted on $\{\bm{u}\}$).~\cite{Waghmare:C:B:2003}
Direct calculation of the forces in real space
requires a computational time proportional to $N^2$, i.e., $O(N^2)$,
where $N$ is the supercell size ($N=L_x\times L_y \times L_z$).
It decreases to $O(N\log N)$ if one calculates the forces in reciprocal space
using fast-Fourier transform (FFT) methods.
This acceleration in computational speed enabled us to perform
MD simulations on $\{\bm{u}\}$ in a large supercell, and was applied
to bulk relaxor ferroelectrics~\cite{Waghmare:C:B:2003,Burton:C:W:PRB:72:p064113:2005}.

Here, we explain how the fast MD method for simulating a first-principles
effective Hamiltonian can be applied to study ferroelectric thin-film
capacitor structures with short-circuited electrodes or external
electric fields.
This new MD method can simulate perovskite-type ferroelectric thin-film capacitors
with dead layers and
consequent depolarization fields.
The high speed of this MD method enables us to simulate
a ferroelectric material
for a realistic system size (up to 100~nm)
and a realistic time span ($>$ 1~ns).

In the next section,
we explain the formalism of the new MD-simulation method for thin-film capacitors.
Results of simulations of BaTiO$_3$ bulk and thin-film capacitors are shown in Sec.~\ref{sec:results}.
In subsection~\ref{subsec:Bulk},
we confirm the reliability of our MD program by simulating thermal properties of bulk BaTiO$_3$.
The advantage of this MD method compared to the Monte-Carlo method is also discussed.
In subsection~\ref{subsec:Capacitors},
we perform heating-up and cooling-down simulations for thin-film BaTiO$_3$ capacitors
with perfect and imperfect electrodes.
Thickness dependence of simulated striped domain structures in thin-film capacitors
with imperfect electrodes are analyzed in detail.
We have already reported some simulated results
of thin-film capacitors of this subsection
and determined thermal properties
in Ref.~[\onlinecite{Paul:N:K:W:PRL:99:p077601:2007}] briefly.
In subsection~\ref{subsec:HysteresisLoops},
newly obtained simulated results of hysteresis loops of thin-film capacitors are reported.
In Sec.~\ref{sec:summary}, we summarize the paper.

We named our MD program {\tt feram} and
distribute it as free software through \url{http://loto.sourceforge.net/feram/}.

\section{Formalism and method of calculation}
\label{sec:Formalism}
\subsection{Effective Hamiltonian}
\label{sub:Effective:Hamiltonian}
The effective Hamiltonian used in the present MD simulations
is basically the same as that in Ref.~[\onlinecite{Waghmare:C:B:2003}].
Here, we present the Hamiltonian
with a notation similar to that in Ref.~[\onlinecite{Zhong:V:R:PRB:v52:p6301:1995}] as
\begin{multline}
  \label{eq:Effective:Hamiltonian}
  H^{\rm eff}
  = \frac{M^*_{\rm dipole}}{2} \sum_{\bm{R},\alpha}\dot{u}_\alpha^2(\bm{R})
  + \frac{M^*_{\rm acoustic}}{2}\sum_{\bm{R},\alpha}\dot{w}_\alpha^2(\bm{R})\\
  + V^{\rm self}(\{\bm{u}\})+V^{\rm dpl}(\{\bm{u}\})+V^{\rm short}(\{\bm{u}\})\\
  + V^{\rm elas,\,homo}(\eta_1,\cdots\!,\eta_6)+V^{\rm elas,\,inho}(\{\bm{w}\})\\
  + V^{\rm coup,\,homo}(\{\bm{u}\}, \eta_1,\cdots\!,\eta_6)+V^{\rm coup,\,inho}(\{\bm{u}\}, \{\bm{w}\})\\
  -Z^*\sum_{\bm{R}}\bm{\mathcal{E}}\!\cdot\!\bm{u}(\bm{R})~,
\end{multline}
where
$\bm{u}=\bm{u}(\bm{R})$ and $\bm{w}=\bm{w}(\bm{R})$
are, respectively,
the local soft-mode amplitude vector and the local acoustic displacement vector
of the unit cell at $\bm{R}$,
the $\alpha$ component of $\bm{R}$ runs over
\begin{equation}
  \label{eq:R}
  R_\alpha = 0,\ a_0,\ 2 a_0,\ \cdots\ (\La-1)a_0~,
\end{equation}
$\eta_1,\cdots\!,\eta_6$ are the homogeneous strain components, and
$M^*_{\rm dipole}$ and $M^*_{\rm acoustic}$ are the effective masses for $\bm{u}$ and $\bm{w}$, respectively.
Note that $\bm{u}$ can also be considered as the optical displacement,
in contrast to
the acoustic displacement $\bm{w}$, or the
dipole moment $Z^*\bm{u}$, where $Z^*$ is the Born effective charge
associated with the soft mode.
In the effective Hamiltonian (\ref{eq:Effective:Hamiltonian}),
external electric field $\bm{\mathcal{E}}$ is taken into account through
its vector product with each dipole moment $Z^*\bm{u}$.

To determine the effective mass $M^*_{\rm dipole}$,
let $\epsilon_\alpha^\tau(\bm{k},i)$ be a mass-weighted $i$-th eigenvector
of the {\em phonon} dynamical matrix at wavevector $\bm{k}$.
Its eigenvalue $\{\omega(\bm{k},i)\}^2$ is the corresponding phonon frequency.
Moreover,
let $d_\alpha^\tau(\bm{k},i)=\epsilon_\alpha^\tau(\bm{k},i)/\sqrt{M_\tau}$
be an atomic displacement vector, which is
normalized as $\sum_{\alpha, \tau} \{d_\alpha^\tau(\bm{k},i)\}^2 = 1$
by adjusting the norm of $\bm{\epsilon}(\bm{k},i)$.
Here, $M^\tau$ is the mass of atom $\tau$.
Generally,
the effective mass of a phonon is $\bm{k}$- and mode-dependent:
\begin{equation}
  \label{eq:EffectiveMassGeneral}
  M^*\!(\bm{k},i) = \sum_{\alpha, \tau} \{d_\alpha^\tau(\bm{k},i)\}^2 M^\tau~.
\end{equation}
However, as an approximation,
we have to employ a unique effective mass for dipoles in the MD simulation.
Thus using the {\em steepest descent} $\Gamma_{15}$ soft-mode normalized
direction vectors
$\bm{\xi}_z = (0.20,\ 0.76, -0.21, -0.21, -0.53)$ and
$\bm{\xi}_x = \bm{\xi}_y = 0$
from Ref.~[\onlinecite{Zhong:V:R:PRB:v52:p6301:1995}], for BaTiO$_3$,
we set $M^*_{\rm dipole}$ as
\begin{equation}
  \label{eq:EffectiveMassXi}
  M^*_{\rm dipole} = \sum_\tau \{\xi_z^\tau\}^2 M^\tau = 39.0\,{\rm amu}~.
\end{equation}
It should be mentioned that $\xi_z^\tau$ is {\em not} equal to
the $d_\alpha^\tau$ of the $\Gamma_{15}$ soft-mode of phonon,
because $M^A$, $M^B$, and $M^{\rm O}$ are not identical.
%

The local-mode self-energy $V^{\rm self}(\{\bm{u}\})$ is
\begin{multline}
  \label{eq:V:self}
  V^{\rm self}(\{\bm{u}\}) =
  \sum_{i=1}^N
  \Bigl\{
  \kappa_2 u^2(\bm{R}_i)
  + \alpha   u^4(\bm{R}_i) + \\
   \gamma
    \left[
      u_y^2(\bm{R}_i) u_z^2(\bm{R}_i) +
      u_z^2(\bm{R}_i) u_x^2(\bm{R}_i) +
      u_x^2(\bm{R}_i) u_y^2(\bm{R}_i)
    \right]
  \Bigr\}~,
\end{multline}
where $u^2(\bm{R}_i)
= u_x^2(\bm{R}_i)
+ u_y^2(\bm{R}_i)
+ u_z^2(\bm{R}_i)$.

The long-range dipole-dipole interaction $V^{\rm dpl}(\{\bm{u}\})$ is
\begin{equation}
  \label{eq:V:dpl}
  V^{\rm dpl}(\{\bm{u}\})=
  \frac{1}{2}\sum_{i=1}^N \sum_\alpha \sum_{j=1}^N \sum_\beta
  u_\alpha(\bm{R}_i) \Phi_\ab(\Rij) u_\beta(\bm{R}_j)~,
\end{equation}
where
\begin{equation}
  \label{eq:Phi}
  \Phi_\ab(\Rij)
  = \frac{Z^{*2}}{\eInf}\sum_{\bm{n}}\!'
  \frac{\dab - 3(\widehat\Rijn)_\alpha(\widehat\Rijn)_\beta}{|\Rij + \bm{n}|^3}~,
\end{equation}
$\eInf$ is the optical dielectric constant (or refractive index squared),
$\dab$ is the Kronecker delta,
a hat indicates a unit vector,
$\bm{n}$ is the supercell lattice vector
\begin{equation}
  \label{eq:n}
  n_\alpha = \cdots,\ -2\La a_0,\ -\La a_0,\ 0,\ \La a_0,\ 2\La a_0,\ \cdots\ \ \ ,
\end{equation}
and $a_0$ is the equilibrium lattice constant.
In Eq.~(\ref{eq:Phi}), $\sum'$ indicates that the summation does not include terms
for which $\Rij=\bm{n}=0$.

We take account of short-range interactions between the optical displacements $\bm{u}(\bm{R})$
up to third nearest neighbor (3nn) as
\begin{equation}
  \label{eq:V:short}
  V^{\rm short}(\{\bm{u}\})=
  \frac{1}{2}\sum_{i=1}^N \sum_\alpha \sum_j^{\rm 3nn} \sum_\beta
  u_\alpha(\bm{R}_i) \,J_{ij,\alpha\beta}\, u_\beta(\bm{R}_j)~,
\end{equation}
where  $J_{ij,\alpha\beta}$ is the short-range interaction matrix,
which can be classified into
7 independent interaction parameters,~\cite{Zhong:V:R:PRB:v52:p6301:1995}
$J_{ij,\alpha\beta} = \pm j_k\ (k=1,\cdots,7)$.

In practice,
$\kappa_2 u_i^2$ in Eq.~(\ref{eq:V:self}),
Eq.~(\ref{eq:V:dpl}), and
Eq.~(\ref{eq:V:short}),
in which $u_\alpha$ is quadratic,
are gathered and calculated in reciprocal space as
\begin{equation}
  \label{eq:V:quad}
  V^{\rm quad}(\{\bm{u}\})=\frac{1}{2} \sum_{\bm{k}} \sum_{\alpha,\beta}
  \widetilde{u}_\alpha^*(\bm{k}) \widetilde\Phi_\ab^{\rm quad}(\bm{k}) \widetilde{u}_\beta(\bm{k}),
\end{equation}
where $\widetilde{u}_\alpha(\bm{k})$ is the Fourier transform
\begin{equation}
  \label{eq:u:FT}
  \widetilde{u}_\alpha(\bm{k}) = \sum_{\bm{R}}u_\alpha(\bm{R})\exp(-i\bm{k}\cdot\bm{R})~,
\end{equation}
of $u_\alpha(\bm{R})$, $\widetilde\Phi_\ab^{\rm quad}(\bm{k})$ is similarly the Fourier transform
of the quadratic interaction matrix (which is only calculated once
at the beginning of simulation\cite{Waghmare:C:B:2003}),
and $\bm{k}$ is a reciprocal vector in the first Brillouin zone of the unit cell such as
\begin{equation}
  \label{eq:k}
  k_\alpha = -\frac{\La-1}{2\La}\Gunit,\ \cdots,\ -\frac{1}{\La}\Gunit,\ 0,\ %
  \frac{1}{\La}\Gunit,\ \cdots,\ \frac{1}{2}\Gunit\ \ \ .
\end{equation}

The homogeneous elastic energy $V^{\rm elas,\,homo}(\eta_1,\cdots\!,\eta_6)$ is
\begin{eqnarray}
  \label{eq:V:elas:homo}
  \nonumber
  V^{\rm elas,\,homo}(\eta_1,\cdots\!,\eta_6)
  & = & \frac{N}{2}B_{11}(\eta_1^2+\eta_2^2+\eta_3^2)\\
  \nonumber
  & + & N          B_{12}(\eta_2\eta_3+\eta_3\eta_1+\eta_1\eta_2)\\
  & + & \frac{N}{2}B_{44}(\eta_4^2+\eta_5^2+\eta_6^2)~,
\end{eqnarray}
where $B_{11}$, $B_{12}$, and $B_{44}$ are the elastic constants expressed in energy unit
($B_{11}=a_0^3C_{11}$, $B_{12}=a_0^3C_{12}$, and $B_{44}=a_0^3C_{44}$).

The inhomogeneous elastic energy $V^{\rm elas,\,inho}(\{\bm{w}\})$
is also calculated in reciprocal space as
\begin{equation}
  \label{eq:V:elas:inho}
  V^{\rm elas,\,inho}(\{\bm{w}\}) = \frac{1}{2} \sum_{\bm{k}} \sum_{\alpha,\beta}
  \widetilde{w}_\alpha^*(\bm{k}) \widetilde\Phi_\ab^{\rm elas,\,inho}(\bm{k}) \widetilde{w}_\beta(\bm{k}).
\end{equation}
For the {\em force constant} matrix $\widetilde\Phi_\ab^{\rm elas,\,inho}(\bm{k})$,
we employed the long-wavelength approximation.
For instance,
the diagonal part is
\begin{equation}
  \label{eq:diagonal}
  \widetilde\Phi_{xx}^{\rm elas,\,inho}(\bm{k}) = \frac{1}{N}
  \left[ k_x^2 B_{11} + k_y^2 B_{44} + k_z^2 B_{44} \right]~,
\end{equation}
and the off-diagonal part is
\begin{equation}
  \label{eq:off-diagonal}
  \widetilde\Phi_{xy}^{\rm elas,\,inho}(\bm{k}) = \frac{1}{N}
  \left[ k_xk_yB_{12} + k_xk_yB_{44} \right]~.
\end{equation}

The coupling between $\{\bm{u}\}$ and homogeneous strain is
the same as that given in Ref.~[\onlinecite{King-Smith:V:1994}], i.e.,
\begin{equation}
  \label{eq:V:coup:homo}
  V^{\rm coup,\,homo}(\{\bm{u}\}, \eta_1,\cdots\!,\eta_6) =
  \frac{1}{2} \sum_{\bm{R}} \sum_{i=1}^6 \sum_{j=1}^6 \eta_i \, C_{ij} \, y_j(\bm{R})~.
\end{equation}
Here,
$y_1(\bm{R})$ = $u_x^2(\bm{R})$,
$y_2(\bm{R})$ = $u_y^2(\bm{R})$,
$y_3(\bm{R})$ = $u_z^2(\bm{R})$,
$y_4(\bm{R})$ = $u_y(\bm{R})u_z(\bm{R})$,
$y_5(\bm{R})$ = $u_z(\bm{R})u_x(\bm{R})$, and
$y_6(\bm{R})$ = $u_x(\bm{R})u_y(\bm{R})$,
\begin{equation}
  \label{eq:Cij}
  {\bf C} = \left(
    \begin{array}{cccccc}
      B_{1xx} & B_{1yy} & B_{1yy} & 0 & 0 & 0 \\
      B_{1yy} & B_{1xx} & B_{1yy} & 0 & 0 & 0 \\
      B_{1yy} & B_{1yy} & B_{1xx} & 0 & 0 & 0 \\
      0 & 0 & 0 & 2B_{4yz} & 0 & 0 \\
      0 & 0 & 0 & 0 & 2B_{4yz} & 0 \\
      0 & 0 & 0 & 0 & 0 & 2B_{4yz} \\
    \end{array}
  \right)~,
\end{equation}
and $B_{1xx}$, $B_{1yy}$, and $B_{4yz}$ are
the coupling coefficients defined in Ref.~[\onlinecite{King-Smith:V:1994}].

The coupling between $\{\bm{u}\}$ and inhomogeneous strain
is also calculated in reciprocal space as
\begin{equation}
  \label{eq:V:coup:inho}
  V^{\rm coup,\,inho}(\{\bm{u}\}, \{\bm{w}\}) =  \frac{1}{2} \sum_{\bm{k}} \sum_{\alpha} \sum_{i=1}^6
  \widetilde{w}_\alpha(\bm{k}) \widetilde{B}_{\alpha i}(\bm{k}) \widetilde{y}_i(\bm{k})~,
\end{equation}
where $\widetilde{w}_\alpha(\bm{k})$ and $\widetilde{y}_i(\bm{k})$ are
the Fourier transforms of $w_\alpha(\bm{R})$ and $y_i(\bm{R})$, respectively.
For the $3\times6$ coupling matrix ${\bf B}(\bm{k})$,
we again employed the long-wavelength approximation
\begin{widetext}
\begin{equation}
  \label{eq:Balphai}
  \widetilde{\bf B}(\bm{k}) = \frac{1}{N}\left(
    \begin{array}{cccccc}
      k_xB_{1xx} & k_xB_{1yy} & k_xB_{1yy} & 0 & 2k_zB_{4yz} & 2k_yB_{4yz} \\
      k_yB_{1yy} & k_yB_{1xx} & k_yB_{1yy} & 2k_zB_{4yz} & 0 & 2k_xB_{4yz} \\
      k_zB_{1yy} & k_zB_{1yy} & k_zB_{1xx} & 2k_yB_{4yz} & 2k_xB_{4yz} & 0 \\
    \end{array}
  \right)~.
\end{equation}
\end{widetext}

In the present MD simulations of BaTiO$_3$,
the parameters from Refs.~[\onlinecite{Zhong:V:R:1994}] and [\onlinecite{Zhong:V:R:PRB:v52:p6301:1995}],
which are determined by first-principles calculations,
are employed.
As mentioned in Refs.~[\onlinecite{Zhong:V:R:1994}] and [\onlinecite{Zhong:V:R:PRB:v52:p6301:1995}],
this parameter set leads to an underestimation of the Curie temperature $T_{\rm C}$.
To correct this underestimation, we follow these references in applying a
negative pressure of $p=-5.0$~GPa in all simulations.

\subsection{Molecular Dynamics}
MD simulations with the effective Hamiltonian of Eq.~(\ref{eq:Effective:Hamiltonian})
are performed in the canonical ensemble
using the Nos\'e-Poincar\'e thermostat.~\cite{Bond:L:L:JComputPhys:151:p114-134:1999}
This simplectic thermostat is so efficient that we can set the
time step to $\Delta t=2$~fs.
In our present simulations,
we thermalize the system for 40,000 time steps,
after which we average the properties for 10,000 time steps.

\begin{figure}
  \centering
  \includegraphics[width=60mm]{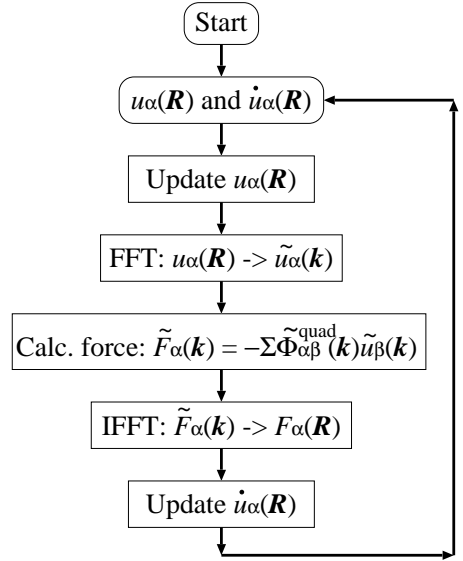}
  \caption{Simplified flow chart for calculating forces on $u_\alpha(\bm{R})$.
     Fast Fourier transform (FFT) and inverse FFT (IFFT) enable
     rapid calculation of long-range dipole-dipole interactions.}
  \label{fig:flow}
\end{figure}

In Fig.~\ref{fig:flow} we roughly illustrate how to calculate
the forces exerted on $u_\alpha(\bm{R})$
with $\widetilde\Phi_\ab^{\rm quad}(\bm{k})$  in Eq.~(\ref{eq:V:quad})
and how the time evolution is simulated.
First, $u_\alpha(\bm{R})$ is FFTed to $\widetilde{u}_\alpha(\bm{k})$,
the force $\widetilde{F}_\alpha(\bm{k}) = -\sum_{\beta} \widetilde\Phi_\ab^{\rm quad}(\bm{k}) \widetilde{u}_\beta(\bm{k})$ is calculated in reciprocal space,
and then the force in real space is obtained
by the inverse FFT of $\widetilde{F}_\alpha(\bm{k})$.
In practice, updates of $u_\alpha(\bm{R})$ and
$\dot{u}_\alpha(\bm{R})=\frac{\partial }{\partial t}u_\alpha(\bm{R})$ are processed
in the manner of the Nos\'e-Poincar\'e thermostat.

The homogeneous strain components $\eta_1,\cdots\!,\eta_6$ are determined
by solving
\begin{multline}
  \label{eq:eta:minimize}
  \frac{\partial}{\partial\eta_i}\Bigl[V^{\rm elas,\,homo}(\eta_1,\cdots\!,\eta_6)\\
+ V^{\rm coup,\,homo}(\{\bm{u}\}, \eta_1,\cdots\!,\eta_6)\Bigr] = 0
\end{multline}
at each time step according to $\{\bm{u}\}$
so that $\eta_1,\cdots\!,\eta_6$ minimize
$V^{\rm elas,\,homo}(\eta_1,\cdots\!,\eta_6) +
 V^{\rm coup,\,homo}(\{\bm{u}\}, \eta_1,\cdots\!,\eta_6)$.
While the local acoustic displacement $w_\alpha(\bm{R})$ could be treated
as dynamical variables using the effective mass $M^*_{\rm acoustic}$,
we have instead chosen to integrate out these variables in a manner
similar to the treatment of the homogeneous strain.  That is,
$w_\alpha(\bm{R})$ is determined
so that $V^{\rm elas,\,inho}(\{\bm{w}\})+V^{\rm coup,\,inho}(\{\bm{u}\}, \{\bm{w}\})$
becomes minimum at each time step according to $u_\alpha(\bm{R})$.
Technically,
the minimization is performed by solving the linear set of equations
\begin{equation}
  \label{eq:w:minimize}
  \widetilde\Phi^{\rm elas,\,inho}(\bm{k}) \widetilde{\bm{w}}(\bm{k}) +
  \widetilde{\bf B}(\bm{k})\widetilde{\bm{y}}(\bm{k})=\bm{0}
\end{equation}
for each $\bm{k}$ in reciprocal space.

\subsection{Ferroelectric Thin Films}
\label{subsec:FerroelectricThinFilm}

\begin{figure}
  \centering
  \includegraphics[width=80mm]{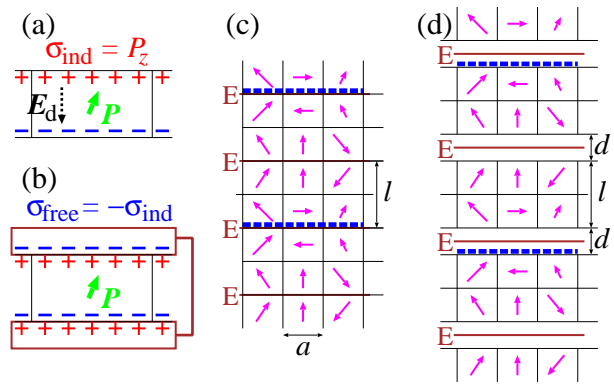}
  \caption{Schematic illustrations of ferroelectric thin films of thickness
    $l$ unit cells (here $l=2$).
    (a)~Isolated thin film in vacuum.
    (b) Thin film sandwiched between short-circuited perfect electrodes.
    Doubly periodic boundary conditions
    for simulations of films sandwiched between
    perfect and imperfect short-circuited electrodes
    are depicted in (c) and (d), respectively.
    Horizontal thick lines marked with ``E''
    represent the electrostatic mirrors used to model electrodes.
    They are a distance $d/2$
    away from the ferroelectric film surface ($d=0$ in~(c), $d=1$ in~(d)).
    Each thin arrow represents a local dipole within
    a unit cell ($a^3=3.94$~\AA$^3$) of the BaTiO$_3$ crystal.
    Thick dashed lines enclose the periodic cell used for simulations.}
  \label{fig:thinfilmes}
\end{figure}

If a ferroelectric thin film is placed in isolation in vacuum
without electrodes as depicted in Fig.~\ref{fig:thinfilmes}(a),
its spontaneous polarization $\bm{P}=(P_x, P_y, P_z)$,
which is represented by a thick arrow in the figure,
induces charges
$\pm\sigma_{\rm ind}=\pm P_z$ at both surfaces,
and the induced charges cause a full depolarization field in the thin film,
$\bm{\mathcal{E}}_{\rm d}=-4\pi\sigma_{\rm ind}\widehat{\bm{z}}
               =-4\pi P_z            \widehat{\bm{z}}$.
On the other hand,
if the ferroelectric thin film is placed between short-circuited
perfect electrodes as depicted in Fig.~\ref{fig:thinfilmes}(b),
the induced charges are fully canceled by
free charges $\sigma_{\rm free}$ arising at both surfaces of the electrodes,
$\bm{\mathcal{E}}_{\rm d}=-4\pi(\sigma_{\rm ind}+\sigma_{\rm free})\widehat{\bm{z}}=0$.
This geometric circumstance can be simulated
with the doubly periodic supercell as depicted in Fig.~\ref{fig:thinfilmes}(c),
because the two electrodes act
as two electrostatic mirrors facing each other,
and the mirrors make oppositely charged infinite mirror images
beyond the electrodes.

We can also introduce dead layers of thickness $d$ between the ferroelectric
thin film and electrodes by constraining the local soft-mode amplitudes to
vanish ($\bm{u}=0$) in these layers, as illustrated
in Fig.~\ref{fig:thinfilmes}(d).
With the dead layers,
the infinite mirror images beyond the electrodes become $\frac{l}{l+d}$ more sparse
than images of the without-dead-layer configuration.
Consequently, the free charges arising at the electrode surfaces
decrease to $\sigma_{\rm free}=-\frac{l}{l+d}\sigma_{\rm ind}$,
where $l$ is the ferroelectric film thickness.
This simulates short-circuited imperfect electrodes
resulting in a depolarization field of
\begin{equation}
  \label{eq:depolarization}
  \bm{\mathcal{E}}_{\rm d} = -4 \pi \frac{d}{l+d} P_z \widehat{\bm{z}}~.
\end{equation}
We can also use a doubly periodic supercell with dead layers for this case.
%
\uline{Physically, the depolarization field of Eq.~(\ref{eq:depolarization})
can arise either from the presence of a dead layer in the ferroelectric
near the interface, or from imperfect screening at the metal electrode,
or both.  We can define an effective screening length for
each of these effects, and we interpret the ``dead-layer thickness'' $d$ of
our model as corresponding to the {\em sum} of these two physical screening lengths.
The screening length associated with the electrode interface appears in
Eq.~(16) of Ref.~[\onlinecite{MEHTA:S:J:JAP:44:p3379-3385:1973}] and
Eq.~(1) of Ref.~[\onlinecite{Dawber:C:L:S:JPhys-CondesMatter:15:pL393-L398:2003}]
and is discussed for the SrRuO$_3$/BaTiO$_3$ interface in
Refs.~[\onlinecite{Sai:K:R:PRB:72:p020101:2005}],
[\onlinecite{Kim:J:K:C:L:Y:S:N:PRL:95:p237602:2005}], and
[\onlinecite{Gerra:T:S:P:PRL:96:p107603:2006}].
Therefore, while the model does not explicitly incorporate
information about the interface screening, this information is
effectively included in the definition of the total
screening length  $d$ in our model.  Thus, for example, simulations
at constant $d$ for various film thicknesses can give the thickness
dependence of the properties of capacitors with a certain interface
structure.}

In the present MD simulations,
the local soft-mode amplitude vectors $\bm{u}$
in dead layers are fixed to zero by the infinitely large mass.
This infinitely-large-mass trick is congenial
to the Nos\'e-Poincar\'e thermostat for maintaining the Nos\'e-Poincar\'e Hamiltonian at zero.
Moreover, this treatment also has another advantage in that the
short-range interactions
between the surfaces of ferroelectric thin film and the electrodes
are automatically truncated.

The depolarization field $\bm{\mathcal{E}}_{\rm d}$ increases the total energy of
the ferroelectric thin film by
$-\bm{P}\cdot\bm{\mathcal{E}}_{\rm d}=4\pi\frac{d}{l+d}P_z^2$.
To avoid forming a depolarization field in ferroelectric thin films,
it is known that the films often develop
striped domain structures.~\cite{Kittel.PhysRev.70.965,Fong:S:S:E:A:F:T:Science:304:p1650-1653:2004,
Bratkovsky:L:PRL:84:p3177-3180:2000,Bratkovsky:L:PRL:87:p179703:2001}
The introduction of the striped domain structure
can eliminate some part of the energy increase $4 \pi \frac{d}{l+d}P_z^2$,
because $P_z$ becomes zero on average.
However, the striped domain structure involves an energy
cost in the short-range interaction $V^{\rm short}(\{\bm{u}\})$,
because it has domain boundaries
between which $\bm{u}$ has opposite direction $\pm z$.
The shorter the wavelength $\lambda$ of the striped domain structure,
the weaker the depolarization field,
but the higher the short-range interaction energy.
The ground state of a ferroelectric thin film
will be decided by a competition between
the long-range dipole-dipole interactions which favor a short-period
domain structure, and domain-wall energy that arises from the
short-range interactions and favors a uniformly polarized structure
or a longer-period striped structure.
In some previous works~\cite{Lai:P:N:K:F:B:S:PRL:96:p137602:2006,
Lai:P:K:B:S:PRB:75:p085412:2007,Lai:P:K:B:S:APL:91:p152909:2007},
the imperfect screening
was mimicked with a parameter.
On the other hand,
our method with doubly periodic boundary condition
does not require any parameters,
because the effect of imperfectness of electrodes
is automatically and implicitly included
in the long-range dipole-dipole interaction $V^{\rm dpl}(\{\bm{u}\})$.

\section{Results and Discussion}
\label{sec:results}
\subsection{Bulk BaTiO$_3$}
\label{subsec:Bulk}
\begin{figure}
  \centering
  \includegraphics[width=75mm]{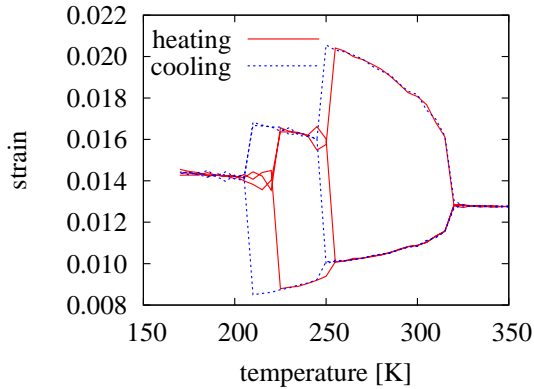}
  \caption{Average homogeneous strains $e_{xx}$, $e_{yy}$, and $e_{zz}$ as a function
  of temperature in heating-up ($+5$~K/simulation, solid lines)
  and cooling-down ($-5$~K/simulation, dashed lines)
  simulations for a $16\times16\times16$ supercell.
  Strains are measured relative to the LDA minimum-energy cubic structure
  with lattice constant 3.948~\AA.}
  \label{fig:strain}
\end{figure}

We first check the reliability of our MD program by comparing results
of our simulations for bulk BaTiO$_3$ with earlier work based on
the same effective Hamiltonian\cite{Zhong:V:R:1994,Zhong:V:R:PRB:v52:p6301:1995}.
We used a system size of $L_x\times L_y\times L_z = 16 \times 16 \times 16$
and small temperature steps in heating-up ($+5$~K/step) and cooling-down
($-5$~K/step) simulations, with initial configuration generated randomly:
$\langle u_\alpha \rangle = 0.07~{\rm \AA}$ and
$\langle u_\alpha^2 \rangle - \langle u_\alpha \rangle^2=(0.02~{\rm \AA})^2$.
We have also checked that there was no dependence of results of these simulations
on initial configurations.
The temperature dependence of the homogeneous strain components (see Fig.~\ref{fig:strain}),
which are the secondary order parameters of ferroelectric phase transitions,
exhibits the correct sequence of phase transitions in BaTiO$_3$ known experimentally.
Even under the negative pressure $p=-5.0$~GPa, the paraelectric to ferroelectric
transition temperature $T_{\rm C}$ is underestimated at around 320~K in comparison with
the experimental value of $T_{\rm C}=408$~K. Our estimates of $T_{\rm C}$'s
agree fairly well with the ones reported in Ref.~[\onlinecite{Zhong:V:R:PRB:v52:p6301:1995}].
The relatively weak first-order nature of the cubic-to-tetragonal phase transition in
comparison with the first-order tetragonal-to-orthorhombic and orthorhombic-to-rhombohedral
phase transitions is evident in the width of the temperature intervals of hysteresis
(see Fig.~\ref{fig:strain}).  We note that the ability to simulate time-dependent phenomena
is one of advantages of MD simulations compared to Monte-Carlo simulations.

\subsection{BaTiO$_3$ ferroelectric thin-film capacitors}
\label{subsec:Capacitors}
We now simulate and analyze the behavior of epitaxially grown films of
BaTiO$_3$ on GdScO$_3$ substrates.~\cite{Choi:B:L:S:S:U:R:C:P:G:C:S:E:Science:306:p1005-1009:2004}
\uline{In our simulations, we represent this with 1\% in-plane biaxial compressive strain
by maintaining the homogeneous strain $\eta_1=\eta_2=-0.01$ and $\eta_6=0$.
In other words, we maintained the {\em average} lattice constants $a$ and $b$ at $0.99a_0$ and angle $\gamma$ at $90^\circ$.}
We use supercell sizes of $L_x\times L_y\times L_z = 32 \times 32 \times 2(l+d)$ and
$40 \times 40 \times 2(l+d)$
and simulate ferroelectric layers of thickness $l$
sandwiched between two short-circuited electrodes
with ($d=1$) and without ($d=0$) dead layers.
This is accomplished through use of doubly periodic boundary conditions as explained earlier.

\begin{figure}
  \centering
  \includegraphics[width=80mm]{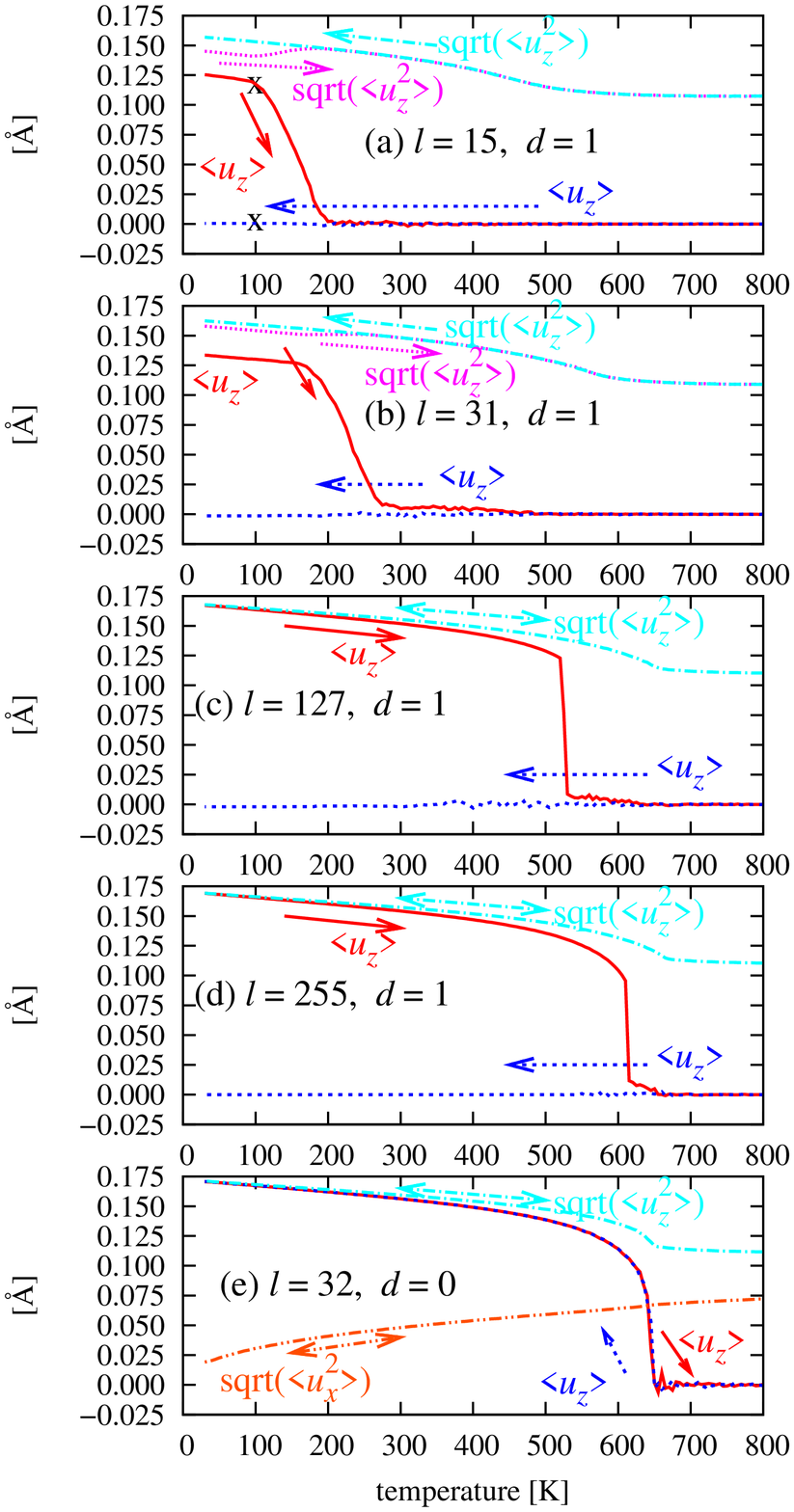}
  \caption{$\langle u_z \rangle$ of
    heating-up (solid lines) and cooling-down (dashed lines)
    molecular-dynamics simulations of
    BaTiO$_3$ thin-film capacitors
    with short-circuited electrodes
    under 1\% in-plane biaxial compressive strain for
    (a) thickness $l=15$~layer with dead layer $d=1$,
    (b) $l=31$  with $d=1$,
    (c) $l=127$ with $d=1$,
    (d) $l=255$ with $d=1$, and
    (e) $l=32$  without dead layer ($d=0$).
    $\sqrt{\langle u_z^2 \rangle}$ are also plotted in (a)-(e).
    In (c)-(e), heating-up $\sqrt{\langle u_z^2 \rangle}$
    and cooling-down $\sqrt{\langle u_z^2 \rangle}$ are almost identical.
    $\sqrt{\langle u_x^2 \rangle}$ is plotted only in (e),
    because the behaviors of
    $\sqrt{\langle u_x^2 \rangle}$ and
    $\sqrt{\langle u_y^2 \rangle}$ are essentially identical in cases (a)-(e)
    (for both heating-up and cooling-down).
    Supercells are of size $32\times 32\times 2(l+d)$.
    Animations of these cooling-down and heating-up simulations are also available
    in the EPAPS\cite{EPAPS}.}
  \label{fig:epit-heat-cool}
\end{figure}

\begin{figure}
  \centering
  \includegraphics[width=88mm]{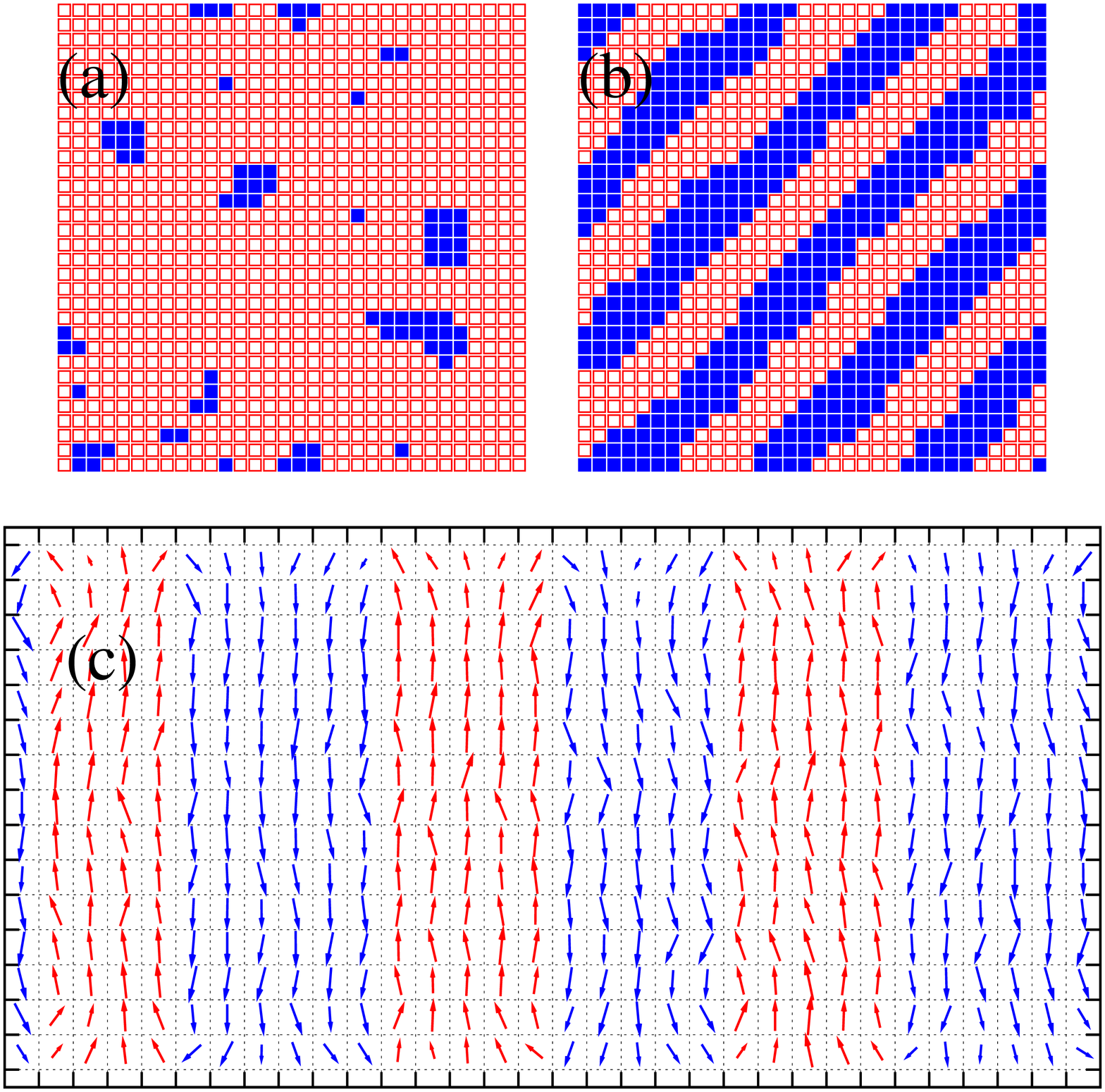}
  \caption{Snapshots at $T=100$~K in
    heating-up   ((a)) and
    cooling-down ((b) and (c)) simulations of ferroelectric
    thin-film capacitors of $l=15$ with $d=1$.
    (a) and (b) are horizontal slices.
    (c) is a vertical cross section.
    Points of snapshots are indicated with ``X'' marks in Fig.~\ref{fig:epit-heat-cool}(a).
    In horizontal slices,
    the $+z$-polarized and $-z$-polarized sites are
    denoted by $\square$ and $\blacksquare$, respectively.
    In vertical cross sections,
    the dipole moments of each site are projected onto the $xz$-plane and indicated with arrows.
    Layers which do not have arrows are dead layers.}
  \label{fig:snapshot}
\end{figure}

Both the heating-up and cooling-down simulations are started with an initial
configuration of
$\langle u_x \rangle = \langle u_y \rangle=0$,
$\langle u_z \rangle=0.07~{\rm \AA}$ and
$\langle u_\alpha^2 \rangle - \langle u_\alpha \rangle^2=(0.02~{\rm \AA})^2$.
In the cooling-down simulations, which start at a sufficiently high temperature,
the initial configuration changes to an unpolarized one ($\langle u_z \rangle=0$)
during thermalization. We monitor the temperature dependence of $\langle u_\alpha \rangle$
and $\langle u_\alpha^2 \rangle$ for thin films with thicknesses $l=15$, 31, 127, and
255 with dead layers $d=1$ and a thin film with thickness $l=32$ without dead layers
($d=0$) (see Fig.~\ref{fig:epit-heat-cool} and animations in the EPAPS\cite{EPAPS}\,). 
The behavior of the film with no dead layer is the same in heating and
cooling simulations.  In contrast, for the films with a dead layer ($d=1$),
the transition behavior exhibited by $\langle u_z \rangle=0$ is rather
different in heating and cooling simulations, although
the temperature dependence of $\langle u_z^2 \rangle$
is almost the same in both the kinds of simulations.
%
%
In the heating-up simulations,
the discontinuity in $\langle u_z \rangle$ as a function of temperature marks a
transition from a ferroelectric state with almost uniform out-of-plane polarization
(Fig.~\ref{fig:snapshot}(a)) to one with a striped domain
structure (Fig.~\ref{fig:snapshot}(b) and (c)). We find that this transition temperature,
$T_{\rm S}(l,d=1)$, exhibits a strong dependence on size $l$.
We note that this transition is missing in the cooling-down simulations;
just above $T_{\rm S}$, striped domain structures appear and the stripes remain and be frozen
at $T<T_{\rm S}$.
The temperature $T_{\rm C}(l,d=1)$ at which
$\sqrt{\langle u_z^2 \rangle}(T)$ exhibits a change in its slope marks another transition,
namely from
a striped domain phase to a paraelectric phase. $T_{\rm C}(l,d=1)$ depends relatively weakly
on the film thickness.
\uline{It should be mentioned that the results of heating-up simulations
with a phase transition from the single domain state to the striped domain state
at $T_{\rm S}$ below $T_{\rm C}$
agree with thermodynamical treatment
of ferroelectric capacitors with dead layers
by Chensky and Tarasenko.~\cite{Chensky:Tarasenko:SovPhysJETP:56:p618:1982}}

For films with $d=1$,
$T_{\rm S}$ is 150 and 210~K for $l=15$ and $l=31$ respectively,
which is lower than the bulk transition temperature ($T_{\rm C}\approx320$~K).
However, for $d=1$ films with $l=127$ and $l=255$, $T_{\rm S}$ is enhanced to 520 and 610 K
respectively, well above the bulk $T_{\rm C}$.
In the infinite thickness limit ($l\rightarrow\infty$),
it appears that $T_{\rm S}(l,d=1)$ tends to the $T_{\rm C}$ of thick films with no dead layer ($d=0$),
since $T_{\rm C}$ is 650 K for $l=32$ and $d=0$.
In the $d=1$ cases with $l=127$ and $l=255$, the effect arising from the depolarization field
weakens significantly, and the enhancement of $T_{\rm S}$ results from
the in-plane biaxial compressive strain.
In the $d=0$ case with $l=32$, there is no depolarization field and enhancement
of $T_{\rm C}$  by the in-plane biaxial compressive strain is effective even in
very thin films. We note that $\sqrt{\langle u_z^2 \rangle}$ and $\sqrt{\langle u_x^2 \rangle}$
are distinct even at high temperatures (see Fig.~\ref{fig:epit-heat-cool}e), indicating that
the symmetry of the paraelectric phase is broken by
the presence of the epitaxial constraint and
the electrodes, as well as correlations between local dipoles and their images.

\begin{figure}
  \centering
  \includegraphics[width=85mm]{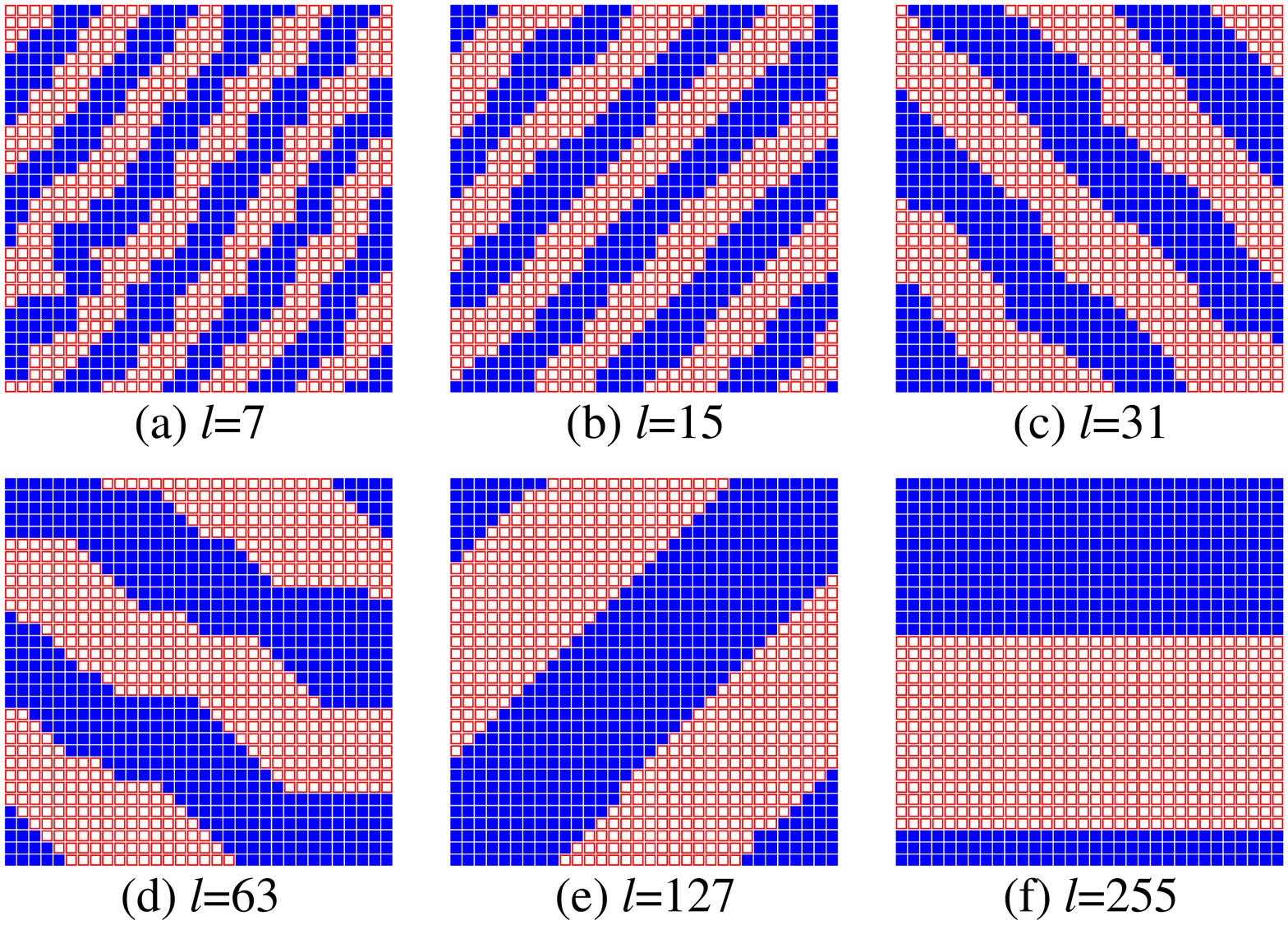}
  \caption{Horizontal slices of snapshots at 100~K
    in cooling-down simulations of ferroelectric
    thin-film capacitors with single dead layer ($d=1$)
    of various thickness $l=7$, 15, 31, 63, 127, and 255.
    The $+z$-polarized and $-z$-polarized sites are
    denoted by $\square$ and $\blacksquare$, respectively.}
  \label{fig:domain}
\end{figure}

\begin{figure}
  \centering
  \includegraphics[width=55mm]{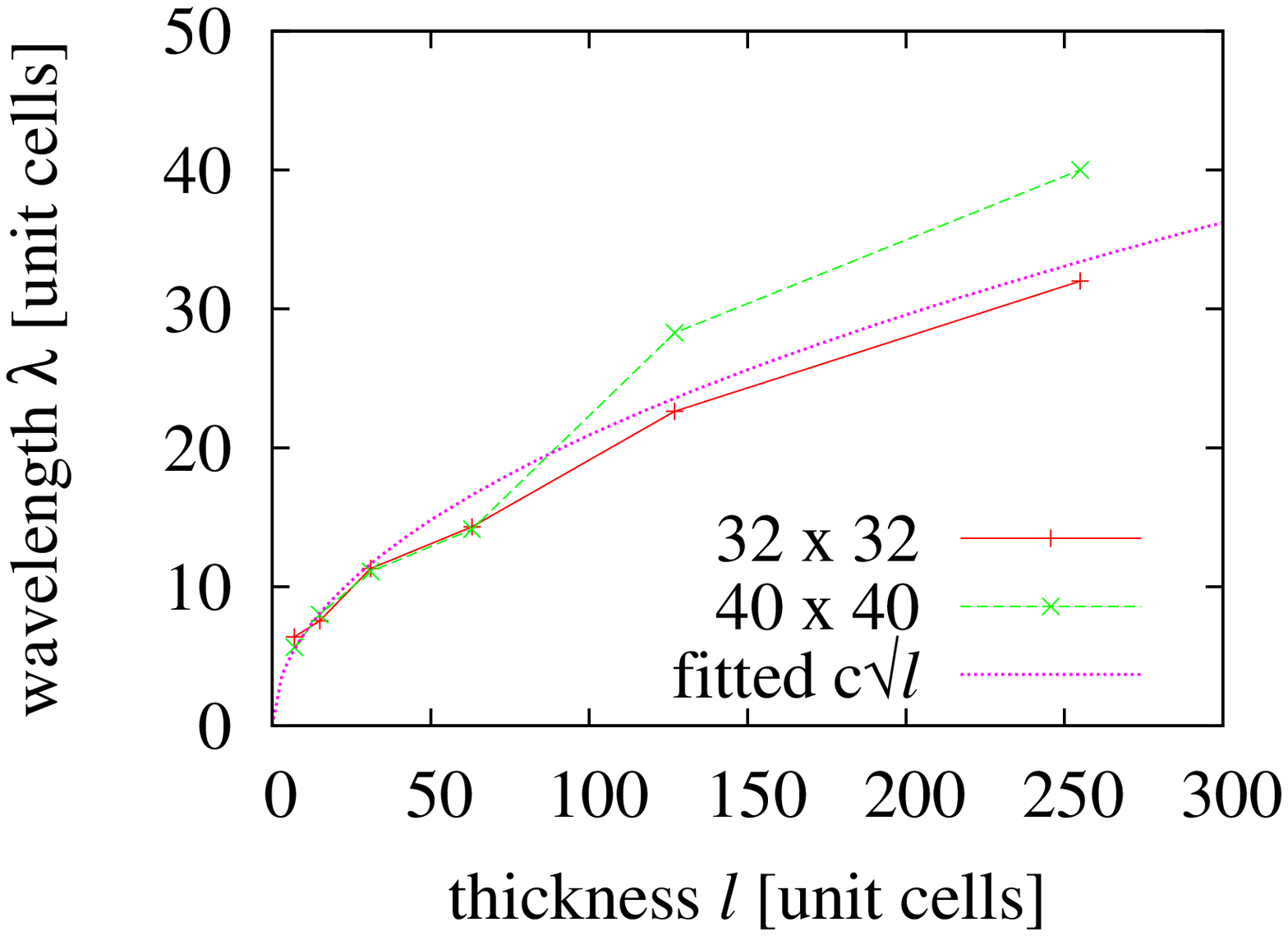}
  \caption{Calculated thickness $l$ dependence of wavelength $\lambda$ of striped
   domain structures in thin film BaTiO$_3$ capacitors with a dead layer ($d=1$).
   $+$ marks are from $32 \times 32 \times 2(l+d)$ supercell
   calculations and $\times$ are those of $40 \times 40 \times 2(l+d)$.
   Data of $l<=127$ are fitted with $\lambda=c\sqrt{l}$ (dotted line). $l=255$ data are
   omitted, because of their large supercell-size dependence.}
  \label{fig:thickness-dep}
\end{figure}

For films with a dead layer ($d=1$), the striped domain structures appear in the
cooling-down simulations at low temperatures for all values of thicknesses $l$ explored here
(see Fig.~\ref{fig:snapshot}(b) and (c) for the case of $l=15$ with $d=1$,
and Fig.~\ref{fig:domain} for various $l$).
As shown in Table~\ref{tab:thickness-dep},
the wavevector $\bm{k}$ of the striped domain,
at which $\widetilde{u}_z(\bm{k})$ has the largest amplitude $|\widetilde{u}_z(\bm{k})|$,
exhibits an interesting dependence on thickness $l$.
We have determined $\bm{k}$ for two supercell sizes, 
$32 \times 32 \times 2(l+d)$ and $40 \times 40 \times 2(l+d)$,
to identify supercell-size effects.
It can be seen that, except for the data for $l=255$,
$\bm{k}$ tends to be along the in-plane $\{ 110 \}$ direction,
consistent with earlier reports.~\cite{Tinte:Stachiotti:PhysRevB.64.235403,Lai:P:K:B:S:PRB:75:p085412:2007}
The simulated striped domain structure for $l=255$,
which is parallel to the $\{ 100 \}$ direction, is likely to be an artifact of the
finite supercell-size:
$L_x\times L_y=32\times 32$ or even $40\times 40$ are too small
to allow for the formation of a sufficiently thick $\{ 110 \}$ striped domain.
\begin{table}
  \caption{Dependence of the wavevector $\bm{k}/2\pi$ of the striped domain structure
    on thickness $l$ in the thin-film BaTiO$_3$ capacitor with a dead layer ($d=1$).}
  \label{tab:thickness-dep}
  \centering
  \begin{tabular}{rrccclrcccl}
    \hline
    $l$ & \multicolumn{5}{c}{$32 \times 32 \times 2(l+d)$} & \multicolumn{5}{c}{$40 \times 40 \times 2(l+d)$} \\
    \hline
      7 & \ \ \ \ \{ & 4/32 & 3/32 & 0 & \} \ \ \ \ \ & \ \ \ \ \{ & 5/40 & 5/40 & 0 & \} \ \ \ \ \ \\
     15 & \ \ \ \ \{ & 3/32 & 3/32 & 0 & \} \ \ \ \ \ & \ \ \ \ \{ & 4/40 & 3/40 & 0 & \} \ \ \ \ \ \\
     31 & \ \ \ \ \{ & 2/32 & 2/32 & 0 & \} \ \ \ \ \ & \ \ \ \ \{ & 3/40 & 2/40 & 0 & \} \ \ \ \ \ \\
     63 & \ \ \ \ \{ & 2/32 & 1/32 & 0 & \} \ \ \ \ \ & \ \ \ \ \{ & 2/40 & 2/40 & 0 & \} \ \ \ \ \ \\
    127 & \ \ \ \ \{ & 1/32 & 1/32 & 0 & \} \ \ \ \ \ & \ \ \ \ \{ & 1/40 & 1/40 & 0 & \} \ \ \ \ \ \\
    255 & \ \ \ \ \{ & 1/32 & 0/32 & 0 & \} \ \ \ \ \ & \ \ \ \ \{ & 1/40 & 0/40 & 0 & \} \ \ \ \ \ \\
    \hline
  \end{tabular}
\end{table}
The wavelength $\lambda=2\pi/|\bm{k}|$ of dominant periodicity of the domain pattern
is shown as a function of thickness $l$ in Fig.~\ref{fig:thickness-dep}, where it is
evident that the thinner films have smaller $\lambda$
to avoid the stronger depolarization field, Eq.~(\ref{eq:depolarization}).
The fitting shown in Fig.~\ref{fig:thickness-dep} suggests a square-root dependence~\cite{Kittel.PhysRev.70.965}
on $l$ (the result for $l=255$ is not included in the fit).
Extensive simulations at larger length scales would probably be required to clarify
further the dependence of the domain period of these striped structures on film thickness
and dead-layer thickness.

\begin{figure}
  \centering
  \includegraphics[width=88mm]{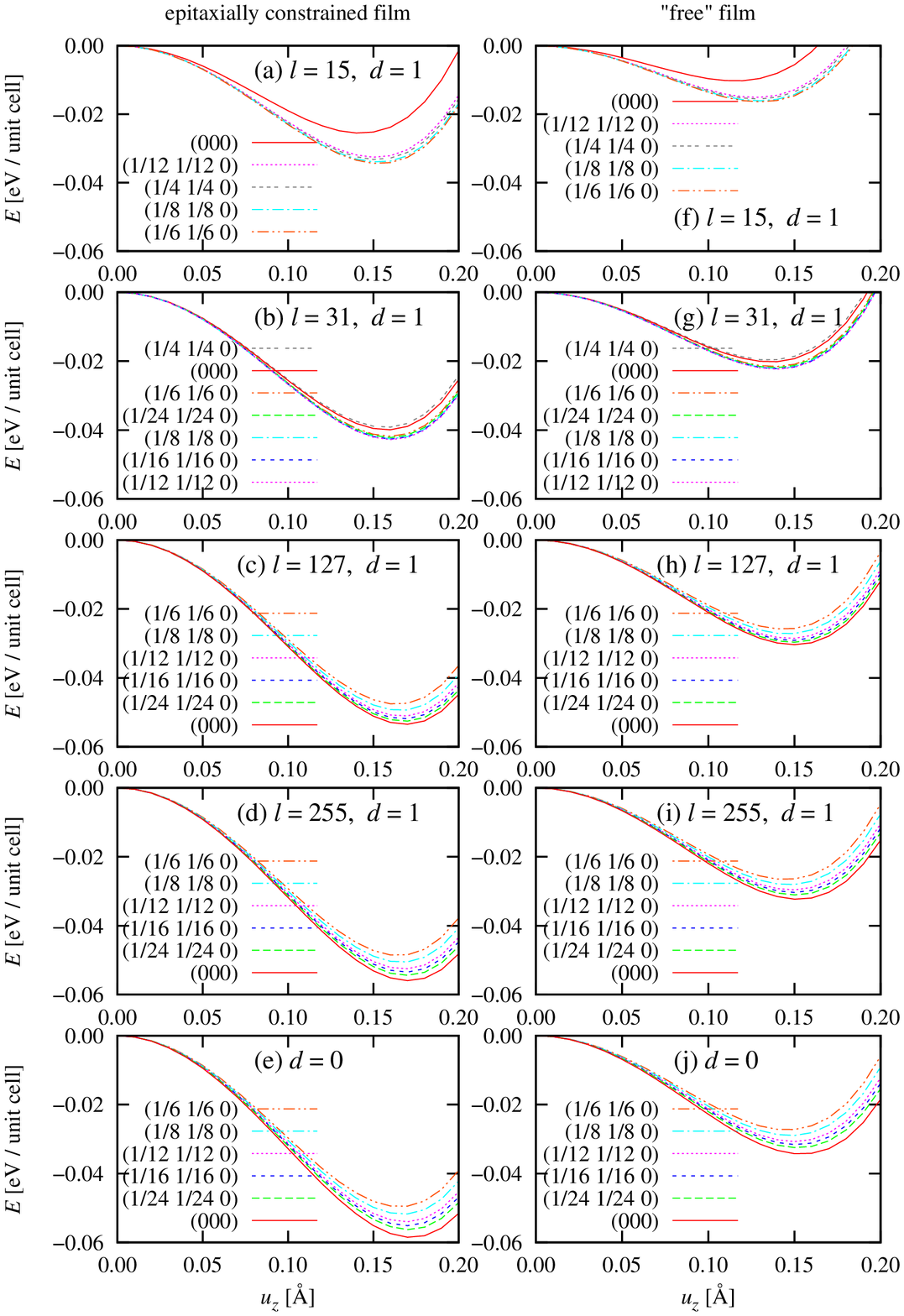}
  \caption{Effective potential surfaces of BaTiO$_3$ thin-film capacitors
    with short-circuited electrodes:
    (a)-(e), under 1\% in-plane biaxial compressive strain arising from epitaxial constraints;
    (f)-(j), without epitaxial constraints (i.e., for ``free'' films).
    The thicknesses of ferroelectric films and dead layers are indicated in each panel
    with $l$ and $d$ respectively.
    Total energies as functions of $u_z$
    are compared among striped domain structures with wavevectors $\bm{k}$ parallel to $(110)$.
    $\bm{k}=(000)$ corresponds to the uniformly polarized structure.
    The zero of the energy scale is placed at the total energy of the non-polarized $u_z=0$ structure.
    A negative pressure $p=-5$~GPa is applied to correct the underestimation in $T_{\rm C}\,$.}
  \label{fig:potential:surface}
\end{figure}

\begin{figure}
  \centering
  \includegraphics[width=55mm]{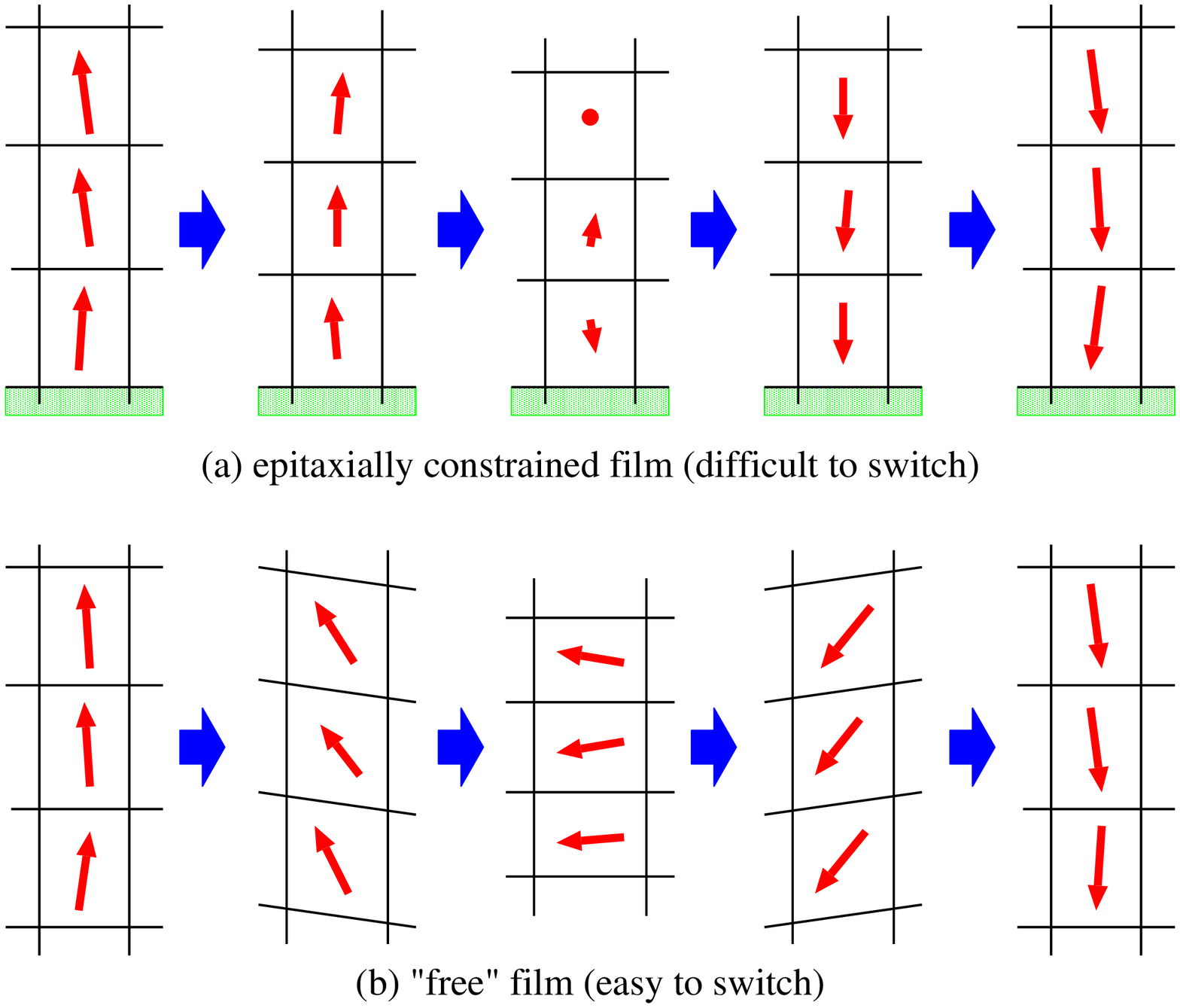}
  \caption{Schematic comparison between the epitaxially constrained film and the ``free'' film.
    In the epitaxially constrained film, switching may have to climb over a potential barrier,
    but, in the ``free'' film,
    dipoles can be easily rotated and
    switching can go around a valley of the potential.}
  \label{fig:epit-vs-free}
\end{figure}

The stark difference in the  behavior of $\langle u_z \rangle$ in heating-up and
cooling-down simulations hints that
the (almost) uniformly polarized state and the
$\langle u_z \rangle=0$ striped domain states
are frozen and
thermal hopping between them may be almost impossible
at low temperatures.
To understand why both uniformly polarized and striped domain states
are stable and thermal hopping between them are difficult,
we investigated the effective potential-energy surfaces for
striped domain structures of various stripe wavevectors $\bm{k}$
and various $l$ for thin-film ferroelectric capacitors with and
without the dead layer (see Fig.~\ref{fig:potential:surface}(a)-(e)).
Omitting surface relaxations in this analysis may be reasonable
because the surface relaxations are confined to the surface region
of the ferroelectric thin films as shown in Fig.~\ref{fig:snapshot}(c).
It can be seen that thinner ferroelectric film have
a shorter stripe wavelength $\lambda=2\pi/|\bm{k}|$
in their ground states. As the thickness $l$ is increased to
$l \approx 127$, the ground state changes from
the striped domain structure to the out-of-plane uniformly-polarized ferroelectric
structure ($\bm{k}=(000)$). However, on the time scale of our simulations ($\approx 1$~ns),
even at $l\approx255$ there is no hopping from the striped domain metastable
state to the uniformly polarized ground state (Fig.~\ref{fig:epit-heat-cool}(d)).
It can also be seen in Fig.~\ref{fig:potential:surface}(a)-(e) that the
magnitude of $u_z$ which gives the minimum-energy ground state becomes larger,
and the minimum energy gets deeper, as $l$ increases, in
good correspondence with the thickness dependence of $T_{\rm C}$.
The trend of $\bm{k}$ with $l$ also shows
good agreement with the simulated values shown in Table~\ref{tab:thickness-dep}.
\uline{The simulated stability of the out-of-plane uniformly-polarized states
against the energetically lower striped-domain states
in thinner ($l<127$) films at low temperature seems
to give support to the recent idea of elastic stabilization of a
homogeneously polarized state in strained ultrathin films.\cite{Pertsev:K:PRL:98:p257603:2007}}
As shown in Fig.~\ref{fig:epit-vs-free}(a), the polarization switching in the epitaxially
constrained film may be suppressed by the presence of a potential barrier that prevents
hopping between the uniformly-polarized and striped-domain states.
For $l \le 127$ with $d=1$, it is expected that a uniformly polarized film would evolve
into a striped domain state, or vice versa, over a sufficiently long time at $T<T_{\rm S}$.
However, the time scale of the evolution might be very much longer than the
present simulation time scale ($\sim$1~ns).
It might also be expected that, in the cooling-down simulations of films with
$l \le 127$ and $d=1$,
the uniformly polarized state is obtained at $T<T_{\rm S}$.
Instead, however, we find that stripes appear.  A close inspection
of the simulations shows that the stripes form slightly above
$T_{\rm S}$, initially in a somewhat disordered fashion, presumably
because such a structure provides a good compromise between
energetic and entropic considerations.  The stripes then get frozen
into place, and become better ordered, as the temperature is reduced
below $T<T_{\rm S}$.
Conversely,
in the case of $d=0$ (Fig.~\ref{fig:epit-heat-cool}(e),
depolarization field $\mathcal{E}_{\rm d}=0$),
the striped domain stricture does not appear during
the heating-up and cooling-down simulations.
This may be because, when $\mathcal{E}_{\rm d}=0$,
there is no reason or chance to form a striped domain structure
even just above $T_{\rm C}$.
At $T_{\rm C}$,
direct phase transition form paraelectric phase to
uniformly polarized ferroelectric phase occurs.
Then, below $T_{\rm C}$,
the system tends to be in its ground state,
the uniformly polarized ferroelectric structure.

\subsection{Hysteresis loops}
\label{subsec:HysteresisLoops}
\begin{figure}
  \centering
  \includegraphics[width=60mm]{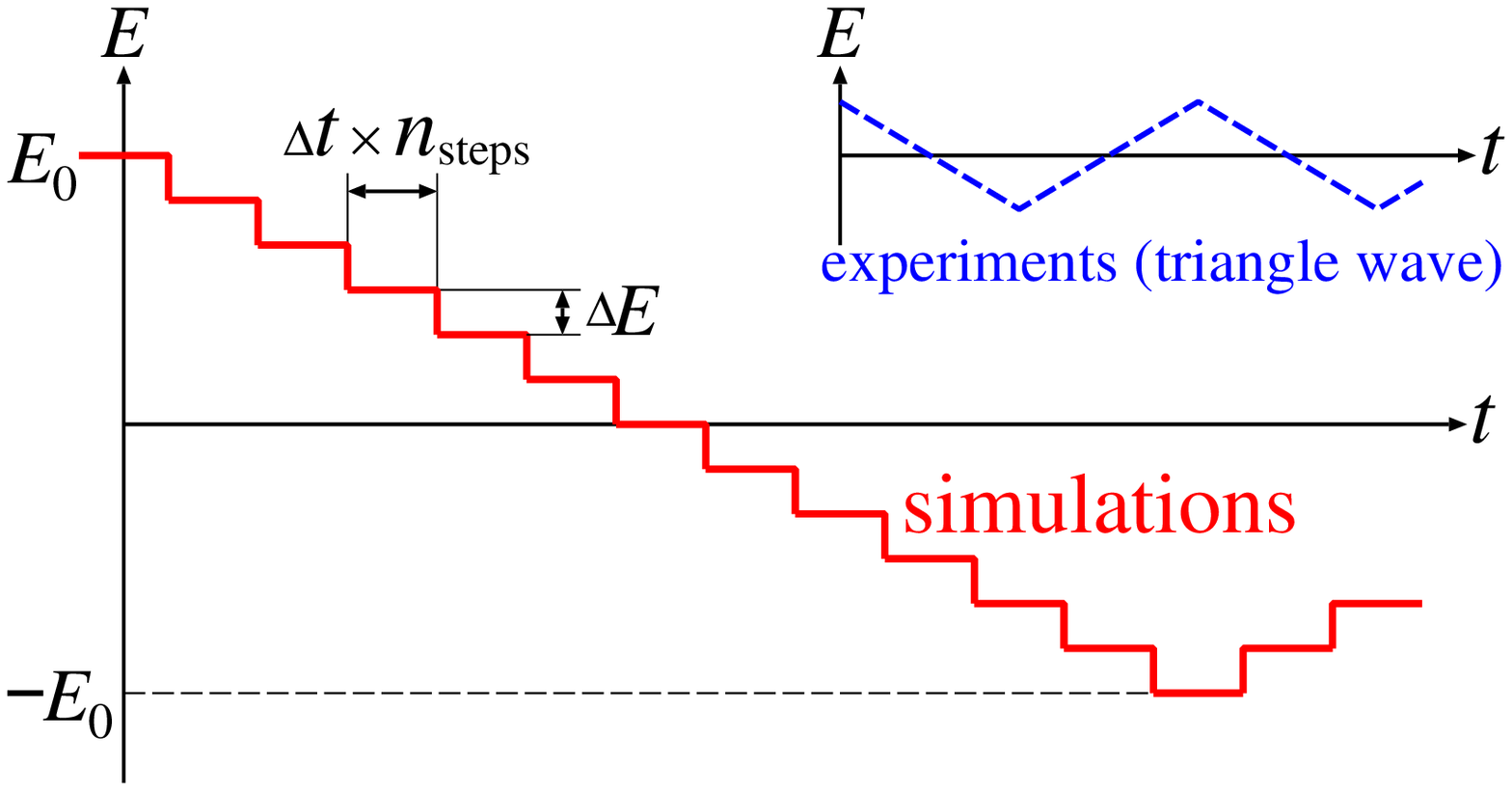}
  \caption{Schematic illustrations of triangle-wave electric field used
    to measure ferroelectric hysteresis loops experimentally (inset)
    and in the present simulations.}
  \label{fig:step}
\end{figure}

\begin{figure}
  \centering
  \includegraphics[width=88mm]{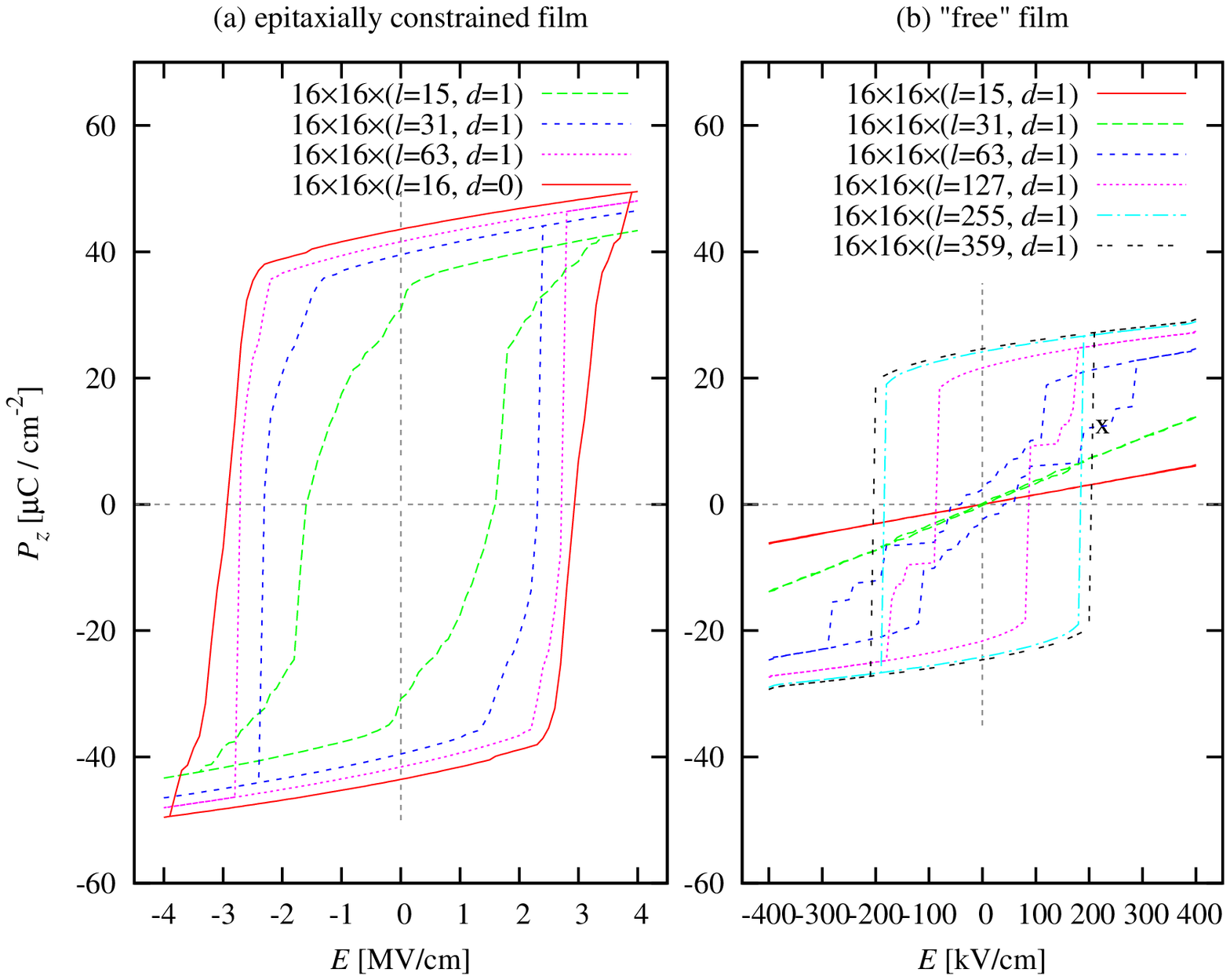}
  \caption{Calculated hysteresis loops for capacitors with
    (a) epitaxially constrained films, and (b) ``free'' films
    of various thickness $l$ and with dead layer $d$.
    Supercell sizes were $16\times 16\times 2(l+d)$.
    $\Delta t=2$~fs and $n_{\rm steps}=$~50,000.
    For epitaxially constrained films,
    $\mathcal{E}_0 =$~4,000~kV/cm and $\Delta \mathcal{E} =$~100~kV/cm are employed.
    For ``free'' films,
    $\mathcal{E}_0 =$~400~kV/cm and $\Delta \mathcal{E} =$~10~kV/cm are employed.}
  \label{fig:epit-vs-free-hysteresis-box}
\end{figure}

\begin{figure}
  \centering
  \includegraphics[width=55mm]{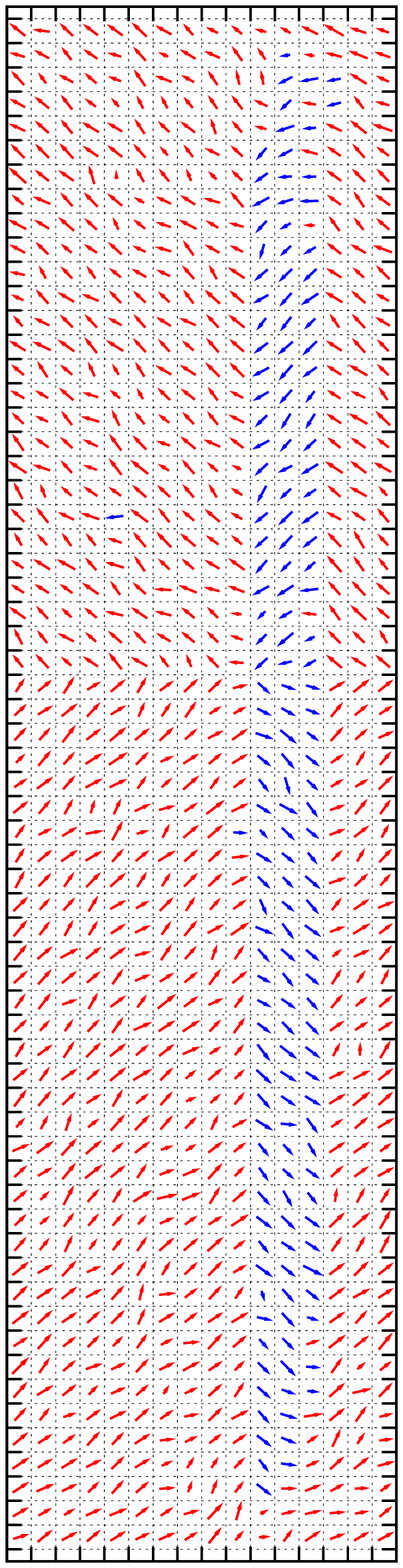}
  \caption{Vertical cross section of a simulated ferroelectric ``free'' film capacitor
    with a single dead layer; $16 \times 16 \times (l=63,~d=1)$.
    The snapshot was taken at the point marked ``x'' in Fig.~\ref{fig:epit-vs-free-hysteresis-box}(b).
    The projection of the dipole moments onto the $xz$-plane are indicated with arrows.}
  \label{fig:damaged}
\end{figure}

A measurement of polarization typically involves use of a
triangle-wave electric field for recording the ferroelectric hysteresis
loops (inset of Fig.~\ref{fig:step}).
The hysteresis loops and coercive fields $\mathcal{E}_{\rm c}$
depend on the amplitude $\mathcal{E}_0$ and frequency $f$ of the applied fields.
We simulate hysteresis here using triangle-wave with steps (width $\Delta t\, n_{\rm steps}$
and height $\Delta \mathcal{E}$) as sketched schematically in Fig.~\ref{fig:step}.
Thus, the frequency of the applied field in our simulations is
$f=\Delta \mathcal{E}/4\, \Delta t\, n_{\rm steps}\mathcal{E}_0$.
We used supercell sizes of $L_x\times L_y\times L_z = 16 \times 16 \times 2(l+d)$ in
simulations of hysteresis loops for ferroelectric thin-film capacitors
with 1\% in-plane biaxial compressive strain and
without constraints of strain (namely, the ``free'' film)
(see Fig.~\ref{fig:epit-vs-free-hysteresis-box}).
\uline{The temperature is maintained at 100~K through the simulations.}
For both the epitaxially constrained and ``free'' films, our simulations confirm
that the imperfect screening of the electrodes decreases the coercive field
as the film thicknesses decreases, as described
phenomenologically in Ref.~[\onlinecite{Dawber:C:L:S:JPhys-CondesMatter:15:pL393-L398:2003}].
There is a large (order-of-magnitude) difference in the
coercive field $\mathcal{E}_{\rm c}$ between
the epitaxially constrained film and the ``free'' film.
This may be because the compressive strain arising from epitaxial constraints
prevents the polarization switching, while the inclusion of
inhomogeneous strain (i.e., acoustic displacements) eases the switching,
as depicted in Fig.~\ref{fig:epit-vs-free}.
The potential barriers themselves are lower in the ``free'' films than
in the epitaxially constrained films (see Fig.~\ref{fig:potential:surface}).
We note that hysteresis loops for ``free'' film capacitors with $l=63$ and $l=127$
are very similar to the experimentally observed hysteresis loops
of a ferroelectric capacitor with damaged electrodes
that have ``steps'' and ``plateaus'' during polarization switchings.~\cite{Scott:Ferroelectric:Memories:2000}
This is because,
in the ``free'' film capacitors with imperfect electrodes ($d=1$),
the configuration with out-of-plane polarization is no longer the ground state.
In fact, the ground state has a nonzero in-plane polarization.
Thus, the dipoles $Z^*\bm{u}(\bm{R})$
have large in-plane components $Z^*u_x(\bm{R})$ and $Z^*u_y(\bm{R})$
in the hysteresis-loop simulations (and experiments), as evident in the snapshot
shown in Fig.~\ref{fig:damaged}.

\uline{Unfortunately, attempts
to fit our results to the usually-assumed Kay-Dunn scaling of the coercive field $\mathcal{E}_{\rm c}$
with film thickness $l$ of thicker 
films\cite{KayDunn1962} were unsuccessful, as were attempts
to emulate the relatively weak dependence of $\mathcal{E}_{\rm c}$
on $l$ for epitaxially grown high-quality ultrathin 
films.\cite{Kim:J:K:C:L:N:S:Y:C:B:K:J:APL:88:p072909:2006,
Jo:K:N:Y:S:APL:89:p232909:2006,Petraru:P:K:P:W:S:K:JAP:101:p114106:2007}
The experimentally observed values of coercive fields $\mathcal{E}_{\rm c}$ for ultrathin BaTiO$_3$ capacitors
range from 200 to 500~kV/cm,\cite{Kim:J:K:C:L:N:S:Y:C:B:K:J:APL:88:p072909:2006,
Jo:K:N:Y:S:APL:89:p232909:2006,Petraru:P:K:P:W:S:K:JAP:101:p114106:2007}
while simulations of epitaxially constrained films largely overestimate $\mathcal{E}_{\rm c}$,
and those of ``free'' films slightly underestimate $\mathcal{E}_{\rm c}$.
This may be because the switching in real thin-film capacitors is a
large-scale ($>$~100~nm) phenomenon involving defect-mediated
nucleation mainly at ferroelectrics-electrodes
interfaces,\cite{KayDunn1962,Janovec1958,
Chandra:coercive:field:Ferroelectrics:313:p7-13:2004,
Kim:J:K:C:L:Y:S:N:PRL:95:p237602:2005,
Tagantsev:G:JAP:100:p051607:2006,
Jo:K:K:C:S:Y:N:PRL:97:p247602:2006}
as well as the possibility that the strain conditions may be intermediate
between the cases of epitaxially constrained and ``free'' films.}
\uline{Such intermediate strain conditions may be achieved and will be simulated with MD in the future
by introducing a mechanical boundary condition such as presented in Ref.~[\onlinecite{Pertsev:K:PRL:98:p257603:2007}].}
\uline{In contrast to our case of ultrathin BaTiO$_3$ capacitors,
it is well known that for ultrathin PbZr$_x$T$_{1-x}$O$_3$ (PZT) capacitors
the coercive fields $\mathcal{E}_{\rm c}$ increase with decreasing
film thickness $l$, and there is an argument whether
this strong increase of $\mathcal{E}_{\rm c}$ is coming from
compressive substrate-induced lattice strain\cite{Pertsev:C:K:H:K:W:APL:83:p3356-3358:2003} or
not.\cite{Lee:N:C:C:R:V:PRL:98:p217602:2007}
Constructing a first-principles Hamiltonian for PZT and simulations with this MD method
will help us to understand this difference between BaTiO$_3$ and PZT.}
%

\section{Summary}
\label{sec:summary}
We have developed a robust and highly efficient molecular-dynamics scheme, based on a
first-principles effective Hamiltonian formulation, for
simulating the behavior of the polarization in perovskite-type ferroelectrics.
We have applied this approach to study BaTiO$_3$ ferroelectric thin-film capacitors,
with special attention to the dependence on film thickness and choice of electric boundary
conditions.  We find that striped domain structures tend to form on cooling-down simulations
when a ferroelectric dead layer is present near the electrodes, and we study the dependence
of the domain period on the conditions of formation.  We also study the hysteresis loops
for capacitor structures, both with and without such dead layers, and we find dramatic
differences in the hysteretic behavior for the cases of elastically constrained or
``free'' films.  Our MD simulator {\tt feram} will be a powerful tool for further
investigations of the physical properties of ferroelectric nanostructures that are
relevant for a variety of potential device applications.

\section*{Acknowledgments}
Computational resources
were provided by the Center for Computational Materials Science,
Institute for Materials Research (CCMS-IMR), Tohoku University.
We thank the staff at CCMS-IMR for their constant effort.
This research was done when T.N. stayed at JNCASR and Rutgers University
under the support from
JNCASR, Rutgers University,
the Ministry of Education, Culture, Sports, Science and Technology (MEXT) of Japan,
and the Japan Society for the Promotion of Science (JSPS).
D.V. acknowledges support of ONR Grant N00014-05-1-0054.

\bibliographystyle{apsrev}
\bibliography{biblio/ferroelectrics,biblio/Bratkovsky-Levanyuk,biblio/FerroelectricCapacitor,biblio/MD}
\end{document}